# The Dynamics of Deterministic Chaos in Numerical Weather Prediction Models

A. Mary Selvam
Deputy Director (Retired)
Indian Institute of Tropical Meteorology, Pune 411 008, India
email: amselvam@eth.net
web sites: http://www.geocities.com/~amselvam
http://amselvam.tripod.com/index.html

## 1. Introduction

Atmospheric weather systems are coherent structures consisting of discrete cloud cells forming patterns of rows/streets, mesoscale clusters (MCC) and spiral bands which maintain their identity for the duration of their appreciable life times in the turbulent shear flow of the planetary Atmospheric Boundary Layer (ABL). The existence of coherent structures (seemingly systematic motion) in turbulent flows has been well established during the last 20 years of research in turbulence. It is still, however debated whether these structures are the consequences of some kind of instabilities (such as shear, or centrifugal instabilities) or whether they are manifestations of some intrinsic universal properties of any turbulent flow (Levich, 1987). The coherent cloud structures in the apparent chaotic (turbulent) flow of the ABL is associated with large values of Reynold's number (Re) up to $10^{12}$ and by convention is described by the inherently non-linear Navier-Stokes (NS) equations. Numerical weather prediction models do not give realistic forecasts because of the following inherent limitations: (1) the non-linear governing equations for atmospheric flows do not have exact analytic solutions and being sensitive to initial conditions give chaotic solutions characteristic of deterministic chaos (2) the governing equations do not incorporate the dynamical interactions and co-existence of the complete spectrum of turbulent fluctuations which form an integral part of the large coherent weather systems (Shafee and Shafee, 1987) (3) limitations of available computer capacity necessitates severe truncation of the governing equations, thereby generating errors of approximations (4) the computer precision related roundoff errors magnify the earlier mentioned uncertainties exponentially with time and the model predictions become unrealistic (Beck and Roepstorff, 1987). The accurate modelling of weather phenomena therefore requires alternative concepts and computational techniques. In this paper a universal theory of deterministic chaos applicable to the formation of coherent weather structures in the ABL is presented with the newly identified computational technique "cellular automata" suitable for computer parallel processing techniques (Hayot, 1987).

## 2. Deterministic Chaos in the ABL

The universal period doubling route to chaos or deterministic chaos is a signature of non-linearity and is found to occur in disparate physical, chemical and biological systems (Feigenbaum, 1980; Fairbairn, 1986; Delbourgo, 1986). Feigenbaum established that the route to chaos is independent of the nonlinear equations describing the system. Lorenz (1963) showed that deterministic chaos is exhibited by the three coupled nonlinear ordinary differential equations for a heat convective system obtained by severe truncation of NS equations. Observational evidence for the existence of deterministic chaos in the ABL has been established conclusively by Lovejoy and Schertzer (1986) who showed that the global cloud cover pattern exhibits fractal geometry which is again a signature of deterministic chaos. Phenomenological observations of fractal (broken or fractured) structure in nature represent the two fundamental symmetries of nature, namely, dilation (r->br) and translation (r->r+c) and correspond respectively to change in unit of length or in the origin of the co-ordinate system (Kadanoff, 1986). A selfsimilar object is identified by its fractal dimension *D* which is defined as *dlnM(R)/dlnR* where *M(R)* is the mass contained within a distance *R* from a typical point in the object. The basic physical mechanism of the observed self-organized fractal geometry in nature is not yet identified (Kadanoff, 1986).

## 3. Physics of Deterministic Chaos in the ABL

The period doubling route to chaos is basically a growth phenomena whereby large eddy growth is initiated from the turbulence scale in successive length step increments equal to the turbulence scale length (Mary Selvam, 1987). In summary, turbulent eddies of frictional origin at the planetary surface possess an inherent upward momentum flux which is progressively amplified by buoyant energy generation from Microscale Fractional Condensation (MFC) of water vapour on hygroscopic nuclei by deliquescence even in an unsaturated environment (Pruppacher and Klett, 1978). The exponential decrease of atmospheric density with height further accelerates the turbulence scale upward momentum flux. Therefore, the unidirectional (upward) turbulence scale energy pump generates successively larger vortex roll circulations in the ABL. The larger eddies carry the turbulent eddies as internal circulations which contribute to their (large eddies) further growth. Such a process of large eddy growth is analogous to the emission of anti-Stokes laser emission triggered by laser pump during chaos in optics (Harrison and Biswas, 1986).

Townsend (1956) has investigated the structure and dynamics of large eddy formations in turbulent shear flows and has shown that large eddies of appreciable intensity form as a chance configuration of the turbulent motion as illustrated in the following example. Consider a large eddy of radius *R* which forms in a field of isotropic turbulence with turbulence length and velocity scales *2r* and *w* respectively. The mean square circulation around a circulation path of large eddy radius *R* is given by

$$W^2 = \oint \oint w w_1 ds\, ds_1$$

where $w$, $w_1$ are the tangential velocity components at the positions of the path elements $ds$ and $ds_1$. If the velocity product falls to zero while the separation $ds$ and $ds_1$ is still small compared with the large eddy radius $R$, i.e. the motion in sufficiently separated paths of flow is statistically independent

$$W^2 = \frac{2}{\pi} \frac{r}{R} w^2 \qquad (1)$$

The above equation can be applied directly to derive the r.m.s. circulation speed $W$ of the large eddy of radius $R$ generated by the turbulence scale energy pump. The scale ratio $Z$ is equal to the ratio of the radii of the large and turbulent eddies. The environment of the turbulent eddy is a region of buoyant energy production by condensation (in the troposphere) and is therefore identified by a Microscale Capping Inversion (MCI) layer on the large eddy envelope. An incremental growth $dR$ of large eddy radius equal to the turbulent eddy radius $r$ occurs in association with an increase $dW$ in large eddy circulation speed as a direct consequence of the buoyant vertical velocity $w_*$ production per second by MFC. The MCI is thus a region of wind shear and temperature inversion in the ABL. The growth of large eddies from the turbulence scale at incremental length steps equal to $r$ - turbulence length scale doubling - is therefore identified as the universal period doubling route to chaos in the ABL.

Therefore, considering turbulence scale yardsticks for length and time, large eddy growth occurs in discrete unit length steps during unit intervals of time and is analogous to 'cellular automata' (Hayot, 1987) computational technique where microscopic domain processes simulate successfully the macroscale flows with simple no scale (scale invariant) analytic equations. The above concept of deterministic chaos is also a 'randomly exact' method of determination of macroscale flow characteristics which is conceived of as a space-time integrated mean of all inherent non-trivial microscopic domain dynamical processes and is therefore scale invariant.

The physics of deterministic chaos therefore enables to identify turbulence as topology dependent and intrinsic to boundary layer flows, the temperature inversion and wind shear being manifestations of large eddy growth from turbulent energy generation processes. This concept is in direct contrast to the conventional view that inversion layers act as atmospheric lids suppressing convective activity and that boundary layer turbulence, e.g. Clear Air Turbulence (CAT) is generated by wind shear in inversion layers, the index of such shear produced turbulence being measured by the Richardson number (Holton, 1979).

## 4. Deterministic Chaos and Atmospheric Eddy Continuum

The turbulent fluctuations mix overlying environmental air into the growing large eddy volume and the steady state non-dimensional fractional volume dilution $k$ of the total large eddy volume across unit cross-section on its envelope is equal to

$$k = \frac{w_*}{dW} \frac{r}{R} \qquad (2)$$

where $w_*$ is the bi-directional turbulent energy acceleration and $dW$ the corresponding acceleration of the large eddy circulation during large eddy incremental length step growth $dR$ equal to $r$. From Eq.(1) it may be computed and shown that $k>0.5$ for $Z<10$. Therefore organized large eddy growth can occur for scale ratios $Z>=10$ only since dilution by environmental mixing is more than half by volume and erases the signature of large eddies for scale ratios $Z<10$. Therefore, a hierarchical, scale invariant, selfsimilar eddy continuum with semi-permanent dominant eddies at successive decadic scale range intervals is generated by the self-organized period doubling route to chaos growth process. The large eddy circulation speed is obtained by integrating Eq.(2) for large eddy growth from the turbulence scale energy pump at the planetary surface and is given as

$$W = \frac{w_*}{k} \ln Z \qquad (3)$$

$k=0.4$ for $Z=10$ from Eqs.(1) and (2). Eq.(3) above is the well known logarithmic wind profile relationship in the surface ABL and $k$ is designated as the Von Karman's constant and its value as determined from observations is equal to *0.4* (Hogstrom, 1985). The deterministic chaos model for eddy dynamics in the ABL therefore predicts that the logarithmic wind profile relationship holds good not only for the surface friction layer, but throughout the ABL. Also, the Von Karman's constant, an arbitrary constant of integration in conventional eddy diffusion theory (Holton, 1979) and determined solely by observations is now shown to have a physical meaning, namely, a no scale (scale invariant) quantification of eddy mass exchange and therefore universal for all hydrodynamic boundary layer fluid flows irrespective of chemical composition and macroscopic size. Von Karman's constant $k$ is therefore more universal than Feigenbaum's constants for deterministic chaos. The Feigenbaum's constants for deterministic chaos are shown in a later section to be functions of $k$ and $Z$.

### 4.1   Semi-permanent dominant eddies (limit cycles) in the ABL

The convective, meso-, synoptic and planetary scale eddies grow from the turbulence scale by the eddy mixing process described above (Section 4.0) at successive decadic scale range intervals. The inherent hierarchy of the atmospheric eddy continuum is manifested as the Meso-scale Cloud Clusters (MCC) in synoptic scale weather systems. The relationship between the radius ($R$), time period ($T$), circulation velocity ($W$) and energy ($E$) scales of the convective (c), meso-(m), synoptic(s) and planetary (p) scale atmospheric systems to the primary turbulence scale ($r$) is derived from Eq.(1) and is given below:

$$R : R_t = r : 10r : 10^2 r : 10^3 r : 10^4 r$$
$$T : T_t = t : 40t : 40^2 t : 40^3 t : 40^4 t \qquad (4)$$
$$E : E_t = \mathcal{E} : 62.5\mathcal{E} : 62.5^2 \mathcal{E} : 62.5^3 \mathcal{E} : 62.5^4 \mathcal{E}$$

The sun is the main source of energy which drives the ABL circulations and therefore periodicities in the weather patterns may finally be related to solar energy input cycles as shown in the following (1) 40-50 day oscillations in the atmospheric general circulation and also the ENSO (≈ 5 years) phenomena may result from the diurnal cycle of solar radiation (1dayx40 ≈ 40-50 days and 1dayx40x40 ≈ 5years) (2) the QBO may result from the semi-diurnal pressure oscillation (12 hoursx40x40 ≈ 2 years) (3) the 22 year oscillation in weather patterns may result from the 5 minute oscillations of the sun's atmosphere (5minsx40$^4$ ≈ 22 years).

A continuous periodogram analysis of high resolution surface pressure values may be used to determine the amplitude and phase of these semi-permanent atmospheric cycles at different locations. Eddy energy enhancement at any scale, for e.g. the $CO_2$ related green house warming effect in the convective scale will result in the total atmospheric eddy continuum energy enhancement (Eq.1) which may be manifested (1) in the synoptic scale features as intensification of small scale short duration intense weather systems such as meso-cyclones and severe local storms. In general seasonal/regional weather anomalies will intensify as a result of tighter coiling of the circulation patterns e.g. the prolonged African drought, erratic monsoon activity and abnormal hurricane tracks. (2) in the planetary scale as an increase in the spatial and temporal domain of the Hadley -Walker circulations with intensification of internal structure as mentioned at (1) above. An extension of temperate rainfall regime to higher latitudes possibly associated with $CO_2$ related global warming has been reported. Further, the ENSO phenomena may also occur with the shorter period of the QBO. Atmospheric eddy continuum energy enhancement due to astronomical causes or human activity is therefore manifested in the first instance as increased variability in global weather patterns leading to perceptible climate change after an appreciable time period. The signature of impending weather/climate change, however can be detected in the geomagnetic H component variations which follow closely changes in atmospheric circulation patterns as shown in a later section.

## 5. Deterministic Chaos and Coherent Helicity

The period doubling route to chaos growth process therefore generates a scale invariant eddy continuum where eddy energy flow structure is in the form of nested logarithmic spiral vortex roll circulations, a complete circulation consisting of the outward and inherent compensating inward flow. The region of chaos is the dynamic growth region of large eddy by turbulence scale energy pumping and the nested vortex hierarchical continuum energy structure is manifested as the strange attractor design with fractal geometry. The atmospheric circulation patterns, therefore have fractal dimensions on all scales ranging from the planetary to the turbulence scale, the strikingly visible pattern of fractal geometry being exhibited by the clouds. The above concept of the steady state turbulent atmospheric boundary layer as a hierarchy of intrinsic helical fluctuations is in agreement with the theoretical investigations of hydrodynamical turbulence by Levich (1987). All basic meso-scale structures (less than 1000km in the tropics) appear to be distinctly helical. These include such outstanding examples of organized geophysical motion as medium scale tornado generating storms, squall lines, hurricanes, etc. (2) Geophysical flows give an implicit indication of the upscale transfer of a certain amount of energy inserted at much smaller scales (3) the helical nature of the most violent geophysical phenomena - a supercell storm - is shown beyond any doubt (Lilly, 1986).

The deterministic chaos model envisages the ABL flow to consist of a web of closed logarithmic spiral circulations anchored to the earth's surface as a unified whole single extended object and having visible manifestation in cloud formation in the troposphere. The atmospheric circulation pattern consisting of dominant eddies at decadic scale range intervals is analogous to (1) the superstrings of a 10-dimensional ($Z=10$) Theory of Everything (TOE) (Fogleman, 1987) (2) the structured quantized vortex roll circulations observed in superfluid Helium (Mineev et al., 1986) and is also similar to self-sustaining solitons or solitary waves, in particular, the triple soliton (Tajima, 1987). The fractal geometry to cloud pattern results from the space-time integration of the non-trivial internal symmetries of the component turbulent eddies and is therefore a manifestation of supersymmetry in nature.

The fractal dimension $D$ of clouds may be expressed as $D=d\ln P/d\ln Z$ or $D=d\ln E/d\ln Z$ where $P$ is the surface pressure, $E$ the eddy kinetic energy and $Z$ the normalised height. Therefore, the spectral slope of the eddy energy spectrum will be equal to the fractal dimension $D$ for the domain $Z$.

The particles in the region of chaos follow laws analogous to Kepler's third law of planetary motion as shown in the following. The periods $T$ and $t$ of the large and turbulent eddies are respectively given as $(2\pi R)/W$ and $(2\pi r)/w$. Substituting for $W/w$ from Eq.(1) gives

$$\frac{R^3}{T^2} = \frac{2}{\pi}\frac{r^3}{t^2}$$

$R^3/T^2$ is a constant for constant turbulence scale energy pump and therefore large eddy circulations follow laws analogous to Kepler's third law of planetary motion. The planetary motions around the sun, the planetary rings around the major planets and the large atmospheric vortices, e.g. polar vortex with structured stratospheric Ozone concentration (Kerr, 1986) may all be manifestations of deterministic chaos.

The rising large eddy gets progressively diluted by vertical mixing due to turbulent eddy fluctuations and a fraction $f$ of surface air reaches the normalised height $Z$ given by

$$f = \frac{W}{w_*}\frac{r}{R}$$

*Therefore* $W = w_* f Z$

From Eqs. (1), (2) and (3)

$$f = \sqrt{\frac{2}{\pi Z}\ln Z}$$

The steady state fractional air mass flux from the surface is dependent only on the dominant turbulent eddy radius.

### 5.1 Atmospheric eddy dynamics independent of Coriolis force

Deterministic chaos model for atmospheric eddy dynamics postulates intrinsic helicity for the solar insolation related major atmospheric circulations as follows. The major planetary scale updraft occurs in the local noon time tropical region with return downdrafts on the local dawn and dusk sectors thereby generating the semi-diurnal pressure oscillation. The air flow into the noon time low is from the west and turning anti-clockwise (clockwise) in the northern (southern) hemispheres because of intrinsic spirality of eddy structures and not due to Coriolis Force as is assumed generally in conventional meteorological theory.

### 6. Deterministic Chaos and Quantum Mechanics

The kinetic energy KE per unit mass of an eddy of frequency $v$ in the hierarchical eddy continuum is shown to be equal to $Hv$ where $H$ is the spin angular momentum of unit mass of the largest eddy in the hierarchy. The circulation speed of the largest eddy in the continuum is equal to the integrated mean of all the inherent turbulent eddy circulations. Let $W_p$ be this mean circulation speed or the zero level about which all the larger frequency fluctuations occur. Therefore

$$KE = \frac{1}{2}W^2 = \frac{1}{2}\frac{2}{\pi}\frac{r}{R}w^2 \qquad \text{from Eq.(1)}$$

and may be written as $KE = HV$

where

$$H = \frac{2}{\pi}\frac{rw^2}{W_p} = R_p W_p$$

$H$ is equal to the product of the momentum of unit mass of planetary scale eddy and its radius and therefore represents the spin angular momentum of unit mass of planetary scale eddy about the eddy center. Therefore the kinetic energy of unit mass of any component eddy of frequency $v$ of the scale invariant continuum is equal to $Hv$. Further, since the large eddy is but the sum total of the smaller scales, the large eddy energy content is equal to the sum of all its individual component eddy energies and therefore the kinetic energy KE distribution is normal and the kinetic energy KE of any eddy of radius $R$ in the eddy continuum expressed as a fraction of the energy content of the largest eddy in the hierarchy will represent the cumulative normal probability density distribution. The eddy continuum energy spectrum is therefore the same as the cumulative normal probability distribution plotted on a log-log scale and the eddy energy probability density distribution is equal to the square of the eddy amplitude. Therefore the atmospheric eddy continuum energy structure follows quantum mechanical laws. The energy manifestation of radiation and other subatomic phenomena appear to possess the dual nature of wave and particles since one complete eddy energy circulation structure is inherently bi-directional and is associated with corresponding bimodal form of manifested phenomena, e.g. formation of clouds in the updraft regions and dissipation of clouds in the downdraft regions giving rise to discrete cellular structure to cloud geometry.

## 7. Deterministic Chaos and Statistical Normal Distribution

The statistical distribution characteristics of natural phenomena follow normal distribution associated conventionally with random chance. The normal distribution is characterized by (1) the moment coefficient of skewness equal to zero, signifying symmetry and (2) the moment coefficient of kurtosis equal to *3* representing intermittency of turbulence on relative time scale. In the following it is shown that the universal period doubling route to chaos growth phenomena in nature gives rise to to the observed statistical normal distribution parameters as a natural consequence. The period doubling route to growth is initiated and sustained by the turbulent (fine scale) eddy acceleration $w_*$ which then propagates by the inherent property of the inertia of the medium. In the context of atmospheric turbulence, the statistical parameters, mean, variance, skewness and kurtosis represent respectively the net vertical velocity, intensity of turbulence, vertical momentum flux and intermittency of turbulence and are given respectively by $w_*$, $w_*^2$, $w_*^3$, $w_*^4$. By analogy, the perturbation speed $w_*$ (motion) per second of the medium sustained by its inertia represents the mass; $w_*^2$, the acceleration (or force); $w_*^3$, the momentum (or potential energy) and $w_*^4$, the spin angular momentum, since an eddy motion is inherently symmetric with bidirectional energy flow, the skewness factor $w_*^3$ is equal to zero for one complete eddy circulation thereby satisfying the law of conservation of momentum. The momentum coefficient of kurtosis which represents the intermittency of turbulence is shown in the following to be equal to *3* as a natural consequence of the growth phenomenon by the period doubling route to chaos. From Eq.(3)

$$dW = \frac{w_*}{k} d(\ln Z)$$

$$\frac{(dW)^4}{w_*^4} = \frac{(d \ln Z)^4}{k^4} = \frac{(dZ)^4}{(kZ)^4}$$

$\frac{(dW)^4}{w_*^4}$ represents the statistical moment coefficient of kurtosis. Organized eddy growth occurs for scale ratio equal to *10* and identifies the large eddy on whose envelope period doubling growth process occurs. Therefore, for a dominant eddy

$$k = \frac{w_*}{W} \frac{r}{R} = \sqrt{\frac{\pi}{22}} \; since \; \frac{r}{R} = \frac{1}{11}$$

*(dZ/Z)=1/2* for one length growth by period doubling process since *Z=dZ+dZ*. Therefore moment coefficient of kurtosis is equal to $\frac{1}{2^4} \times 49 \cong 3$. In other words, period doubling growth phenomena result in a threefold increase in the spin angular momentum of the large eddy for each period doubling sequence. This result is consistent since period doubling at constant pump frequency involves eddy length step growth *dR* on either side of the primary turbulent eddy length *dR*.

## 8. The Universal Feigenbaum's Constants for the ABL

The universal period doubling route to chaos has been studied extensively by mathematicians, the basic example with the potential to display the main features of the erratic behaviour is the Julia model (Delbourgo, 1986) given below.

$$X_{n+1} = F(X_n) = LX_n(1 - X_n)$$

The above nonlinear model represents the population values of the parameter $X$ at different time periods $n$ and $L$ parameterises the rate of growth of $X$ for small $X$. The Eq.(1) representing large eddy growth as integrated space-time mean of turbulent eddy fluctuation given as $W^2 = \frac{2}{\pi}\frac{r}{R}w^2$ is analogous to the Julia model since large eddy growth is dependent on the energy input from the turbulence scale with ordered two-way energy feedback between the larger and smaller scales. Feigenbaum's (1980) research showed that the successive spacing ratios of $X$ and $L$ for adjoining period doublings are given respectively by the two universal constants $a=-2.5029$ and $d=4.6692$. The universal constants $a$ and $d$ assume different numerical values for period tripling, quadrupling etc. and the appropriate values are computed by Delbourgo (1986) and shown to follow the relation $3d=2a^2$ over a wide domain.

The physical concept of large eddy growth by the period doubling process enables to derive the universal constants $a$ and $d$ and their mutual relationship as functions inherent to the scale invariant eddy energy structure as follows.

From Eq.(1) the function $a$ may be defined as

$$a^2 = \left(\frac{WZ}{w_*}\right)^2 = \frac{2Z}{\pi} \cong \frac{2Z}{3} \quad (5)$$

$a$ is therefore equal to $1/k$ from Eq.(2) where $k$ is the Von Karman's constant representing the non-dimensional steady state fractional volume dilution rate of large eddy by turbulent eddy fluctuations across unit cross-section on the large eddy envelope. Therefore '$a$' represents the non-dimensional total fractional mass dispersion rate and is inherently negative and $2a^2$ represents the bi-directional fractional energy flux into the large eddy environment. Let $d$ represent the ratio of the spin angular moments for the total mass of the large and turbulent eddies.

$$\frac{W^4}{w_*^4}Z^3 = \frac{4Z}{\pi^2} \cong \frac{4Z}{9} \quad (6)$$

Therefore $2a^2 = \pi d$ or $2a^2 \cong 3d$ from Eqs.(5) and (6). The above equation relating the universal constants is a statement of the law of conservation of energy, i.e. the period doubling growth process generates a threefold increase in the spin angular momentum of the resulting large eddy and propagates outward as the total large eddy energy flux in the medium. In an earlier section (Section 7) it was shown that the spin angular momentum of the resulting large eddy accounts for the observed value of three for the moment coefficient of kurtosis of the normal distribution. The property of inertia enables propagation of turbulence scale perturbation in the medium by release of the latent energy potential of the medium. An illustrative example is the buoyant energy generation by water vapour condensation in the updraft regions in the ABL.

The universal Feigenbaum's constants *a* and *d* are respectively equal to *-2.52* and *4.05* as computed from Eqs.(5) and (6) since the scale ratio *Z* is equal to *10* for the self-organized eddy growth mechanism in the ABL.

## 9. Deterministic Chaos Model of Weather Systems

The atmospheric weather systems are the visible manifestation of the unified atmospheric eddy continuum in climatologically favourable regions of enhanced buoyant energy generation. The hierarchical helicity inherent to the turbulent shear flow of the ABL is manifested as the Mesoscale Cloud Clusters (MCC) in global weather systems, the cloud bands having inherent curvature (helicity) as exhibited in the strikingly spiral hurricane cloud bands. The deterministic chaos model prediction of the universal and unique patterns of cloud bands and pressure and wind anomaly patterns for synoptic scale weather systems are compared with well documented observational results for the hurricane system (Mary Selvam, 1986). The model concepts are given in the following.

Since large eddy growth involves increase in radius simultaneous with angular displacement from origin, the trajectory of airflow associated with the large eddies will follow a logarithmic spiral pattern both in the horizontal and vertical. The complete eddy circulation consisting of the ascent and the return descent airflow therefore occurs in the form of logarithmic spiral vortices. The full continuum of atmospheric eddies exist as a unified whole in the form of vortices within vortices as displayed in the extreme cases of the tornado funnel and the dust devil. Large eddy growth is initiated at a single point location and growth occurs in a spiral wave form analogous to the self-organized Belousov-Zabotinsky (Ananthakrishnan, 1986) reaction in chemical systems. In the following, quantitative relations are derived for cloud parameters as simple analytic equations from considerations of the microscopic scale dynamics. The angular rotation $d\bar{\theta}$ and the associated incremental radial growth $d\bar{R}$ per second at any location ($R$, $\theta$) is given as

$$d\bar{\theta} = \frac{W}{R} = \frac{w_* f}{r_R}$$

$$d\bar{R} = W = \frac{w_* f R}{r_{R+dR}}$$

$r_{R+dR}$ denotes the turbulent eddy radius corresponding to large eddy of radius *(R+dR)*. The corresponding angle $\alpha$ between the spiral air flow track of the large eddy and the circle drawn with radius *R* is given by

$$\tan\alpha = \frac{dR}{Rd\theta}$$

substituting *b=tan$\alpha$* and integrating for eddy growth from *r* to *R*, the above equation gives

$$R = re^{b\theta}$$

This is the equation for an equiangular logarithmic spiral when the crossing angle $\alpha$ is a constant. At any location A, the horizontal air flow path into the eddy continuum system follows a logarithmic spiral track.

## 9.1 Storm intensity and cloud band configuration

The cloud bands identify the circulation path of the synoptic cyclonic eddy whose radial growth $dR$ is equal to the dominant turbulent eddy radius and $d\theta$ is the corresponding angular rotation.

$$dR = r \text{ and } R = \sum r$$
$$d\theta = t\,d\overline{\theta} \text{ and } \theta = \sum f$$
$$\text{cloud band width} = R\,d\theta = r\,f$$

The dominant turbulent eddy radius determines the angular turning $d\theta$ and incremental large eddy growth $dR$ and therefore the synoptic scale spiral cloud band has different crossing angles and band widths at different locations, with respect to the storm center. Observations show that increased condensation results in decrease in dominant turbulent eddy radius. There is heavy condensation close to the storm center in association with tighter coiling of the spiral with overlapping cloud bands.

## 9.2 Large eddy growth time

The eddy growth time $T$ for an eddy of radius $R$ is computed as follows.

$$T = \frac{dR}{W} = \frac{r}{w_*} \frac{\pi}{2} \left(li\sqrt{Z}\right)_2^z$$

where $li$ is the logarithm integral or Soldner's integral.

## 9.3 Horizontal profile of hurricane pressure field

The low pressure field of the cyclone system is created by the upward ascent of surface air. At any location distance $R$ from the storm center there is an upward mass flux of air equal to $w_*r$ per second across unit area where $r$ is the air density and $w_*$ is the buoyant vertical velocity generated per second by Microscale Fractional Condensation (MFC) at surface layers. A synoptic scale weather system which has been in existence for a time period $T_N$ and extending to a radial distance $R_N$ develops a central pressure departure equal $w_*rT_N$ with respect to the ambient pressure field at the periphery (X). At any intermediate location (say B) the corresponding pressure departure is equal to $w_*rT_R$ where $T_R$ is the time period for the eddy to grow from B to X. The Normalised Pressure Departure (NPD) at the intermediate location with respect to the extreme pressure gradient at the storm center is computed as

$$NPD = \frac{w_* \rho T_R}{w_* \rho T_N} = \frac{T_R}{T_N}$$

### 9.4 Horizontal profile of wind

The horizontal profile of wind ($W$) in a cyclone system follows the logarithmic law and depends only on the turbulent eddy radius from Eq.(3). The airflow speed is mainly due to the dynamic buoyant energy production by MFC and thus is not influenced by the rotation of the earth. Therefore the Coriolis force does not influence the airflow into the synoptic scale eddy as explained earlier. The universal and unique pressure and wind anomaly patterns for the hurricane system is in agreement with reported observations of Holland (1980), Simpson and Riehl (1981) and others.

### 9.5 Quantum mechanics and atmospheric weather systems

The quantum reality which underlies the real world may now be pictured in the context of the universal theory of chaos as applied to the macroscale cloud/weather systems in the ABL and may possibly provide physically consistent commonplace solutions for the apparent inconsistencies and paradoxes (Herbert, 1987) of the quantum mechanics as follows. The unified atmospheric eddy continuum with its complete helical vortex roll circulation consists of balanced and instantaneously adjusting high and low pressure areas of cloud dissipation and formation respectively and may therefore provide the physical analogue for (1) wave-particle dulity (2) non-locality-Berry's phase (3) ultra metric space in sub-atomic particle dynamics.

### 9.6 Cloud dynamics, microphysics and electrification

Cloud growth occurs in the updraft regions of large eddy circulations under favourable conditions of moisture supply in the environment. The turbulent eddies inherent to the large eddies are amplified inside the clouds due to enhanced cloud water condensation and form "cloud top gravity (buoyancy) oscillations". The cloud top gravity oscillations are responsible for (1) cloud vertical mixing and dilution. Downward transport of stratospheric ozone may also occur in deep convective systems as reported by several workers (2) cloud electrification by downward transport of naturally occurring negative space charges from above cloud top regions to the cloud base and simultaneously the upward transport of positive space charges from lower troposphere to the cloud top regions thereby generating the observed vertical positive dipole cloud charge. The Travelling Ionospheric Disturbances (TIDS) have been attributed to cloud top gravity oscillations in deep convective systems. The cauliflower-like surface granularity to the cumulus cloud is a signature of the innumerable turbulent eddies which form the cloud top gravity oscillations.

The deterministic chaos model enables universal no scale (scale invariant) quantification of the steady state cloud dynamical, microphysical and electrical processes (Mary Selvam and Murty, 1985) as listed in the following. (1) The ratio of the actual cloud liquid water content ($q$) to the adiabatic liquid water content ($q_a$) is equal to $f$, the fraction of surface air which reaches the normalised height $Z$ after dilution by vertical mixing due to turbulent eddy fluctuations (2) the vertical profiles of the vertical velocity $W$ and the total cloud liquid water content $q_t$ are respectively given by $W=w_*fZ$ and $q_t=q_*fZ$ where $t$ represents the total values and

\* represents cloud base values (3) the cloud growth time $T = li\sqrt{Z}_2^Z$ where *li* is the logarithm integral (4) the cloud dropsize spectrum follows the naturally occurring Junge aerosol size spectrum and (5) the computed raindrop size spectrum closely resembles the observed Marshall-Palmer raindrop size distribution at the surface (6) the electric field at the surface due to the cloud dipole charge, the strength of the cloud dipole, the cloud electrical conductivity, the point discharge current are expressed in terms of the basic non-dimensional parameters *f* and *Z*. The above quantitative relations are universal for all clouds and depends only on the scale ratio *Z*.

## 10. Deterministic Chaos, Atmospheric Electric Field and Geomagnetic Field

Numerous studies indicate significant correlation between geomagnetic field variations and tropospheric weather activity (Herman and Goldberg, 1978; Gribbins, 1981; Courtillot *et al*., 1982; Kalinin and Rozanova, 1984; Taylor, 1986). However, the exact physical mechanism for the observed coupling between meteorological and geomagnetic phenomena is not yet identified. It may be shown that the atmospheric electric field and geomagnetic field variations are manifestations of the vertical mass exchange process between the lower troposphere and ionosphere. The vertical mass exchange gives rise to upward transport of naturally occurring positive space charges from surface layers and simultaneous downward transport of negative space charges from higher levels. The eddy circulations therefore generate a large scale vertical aerosol current which is of the correct sign and magnitude to generate the horizontal component of the geomagnetic field (Mary Selvam, 1987). Therefore atmospheric circulation patterns leave signature on the geomagnetic field lines whose global variations can be easily monitored by satellite borne sensors and thus assist in weather and climate prediction.

## Conclusion

The deterministic chaos model for atmospheric weather systems enables to formulate governing equations for pressure and wind anomaly patterns in terms of no scale (scale invariant) quantities by consideration of microscopic scale dynamical processes and therefore is analogous to 'cellular automata' computational technique. Parallel processing computers or microprocessors at different locations can be used to compute real time horizontal and vertical profiles of meteorological parameters for global weather systems.